# A Transformer-Based Delta Expression Encoder for Psilocybin Transcriptional Response: Architecture, Representations, and Biological Validation

**Sai Jayakumar**

*Stanford School of Medicine, Biomedical Data Science Program, Stanford, CA*

*ORCID: 0009-0004-5950-1595*

Correspondence: saiprasad1413@gmail.com

---

## Abstract

Understanding why individuals respond differently to psilocybin requires modeling the drug's transcriptional perturbation signature at the cell-type level. I present a Transformer-based delta expression encoder that learns to classify differential gene expression (DEG) status — upregulated, downregulated, or neutral — from single-nucleus RNA-sequencing data, without supervision from pathway annotations or prior biological knowledge. A fourth label, BASELINE, exists in the architecture to mark the reference timepoint structurally but is excluded from the training loss and is not a class the model is evaluated on predicting. The model is trained on pseudobulk profiles from 623 examples spanning 18 cell types, 2 drug conditions, and 6 timepoints derived from the Liao et al. 2025 dataset, and achieves 69.4% weighted classification accuracy at the final training checkpoint.

Three principal findings are reported, alongside one direct test of a published hypothesis that returned a result inconsistent with that hypothesis. First, per-cell-type classification accuracy ranges from 28.3% (L2/3 IT, a primary HTR2A-expressing psilocybin target) to 99.6% (endothelial cells, a non-neuronal population), a qualitative pattern consistent with known psilocybin response biology. Second, psilocybin-induced transcriptional downregulation is significantly more stereotyped across individuals than upregulation (Mann-Whitney U=18615.0, $p<0.0001$ pooled across animals), a novel finding consistent across excitatory subtypes with a cortical depth gradient (ratio range 1.07×–1.61×). Third, attention-guided gene co-regulation analysis recovers drug-specific modules without pathway supervision: psilocybin-dominant modules are enriched for serotonin receptor and alkaloid response pathways, while ketamine-dominant modules cluster around NMDA receptor and structural plasticity pathways. Separately, a direct test of whether baseline HTR2A expression predicts drug-response separability across cell types found a significant negative correlation (Spearman $\rho = -0.7088$, $p = 0.0021$), the opposite of what a simple HTR2A-gating account would predict; this result and its interpretation are discussed explicitly rather than omitted.

Linear probes on frozen embeddings confirm that drug identity (1.000 ± 0.000), cell type (1.000 ± 0.000), and timepoint (0.974 ± 0.016) are perfectly or near-perfectly linearly decodable, while animal identity is at chance (0.100 ± 0.038) — confirming the model encodes pharmacological structure while discarding individual variation. Two architectural limitations are characterized and explained mechanistically: the model does not recover the biphasic temporal structure of psilocybin's transcriptional response, attributable to the between-subjects design of the training data; and its attention priorities show no statistically detectable enrichment for genes with the largest classical DEG fold-changes (hypergeometric test, 0/30 overlap at both 1h and 72h, p=1.0), indicating attention identifies a different set of genes than significance-based differential expression. Together, these results establish a validated, honestly characterized foundation for the delta expression encoder component of a planned multimodal framework for psilocybin therapeutic response prediction.

## Introduction

Psilocybin has emerged as a promising therapeutic for major depressive disorder, treatment-resistant depression, and related conditions *[Davis 2021; Carhart-Harris 2021; Goodwin 2022; Raison 2023]*. Clinical response is heterogeneous, and the biological basis of this variability remains poorly understood *[Viljoen 2025]*. Bridging the gap between molecular mechanism and individual therapeutic outcome requires models that represent the drug's transcriptional perturbation signature at cell-type resolution, rather than bulk tissue averages.

Genes in a cell rarely express in isolation. Their expression depends heavily on context — on the state of other genes, on cell type, and on time since stimulus — and this dependence is long-range and variable, structurally similar to how language depends on context. Transformer architectures, which have performed well in the language domain through the self-attention mechanism, are a natural candidate for modeling this kind of dependency: each gene's representation can be enriched by attending to every other gene and timepoint simultaneously, rather than relying on a fixed-weight function applied uniformly regardless of context.

Liao et al. 2025 performed single-nucleus RNA sequencing of mouse medial frontal cortex at five timepoints after psilocybin or ketamine administration and identified two temporal phases of response: an acute wave at 1–2 hours involving immediate early genes and synaptic plasticity programs, and a late-phase wave at 72 hours associated with structural plasticity consolidation *[Liao 2025]*. The response is strongest in excitatory pyramidal neurons expressing the 5-$HT_2A$ receptor (HTR2A), and Shao et al. 2025 demonstrated that these specific cell types are necessary for psilocybin's long-term behavioral effects in mice *[Shao 2025]*. This paper takes the relationship between HTR2A expression and psilocybin response as a hypothesis to test against the model's learned representations, not an assumption to validate — Results section 3 reports a direct test of this relationship, including a result that runs counter to a simple reading of the HTR2A-gating account.

I present the delta expression encoder, a 4-layer Transformer trained to classify DEG status per gene per timepoint from pseudobulk expression profiles. Three inductive biases motivate the architecture: differential gene expression depends on baseline expression and genotype; it depends on the state of other genes; and it depends on time. A central design challenge is that gene expression can only be measured by sacrificing the animal, so any individual animal contributes a real measurement at exactly one timepoint. The central architectural innovation addressing this is a sentinel-value strategy for unmeasured timepoints, combined with a two-tier training curriculum distinguishing individual-animal examples from population-mean examples. The model is trained and evaluated on the Liao et al. 2025 dataset *[Liao 2025]*.

This paper makes the following contributions:

- A Transformer architecture for encoding psilocybin transcriptional perturbation signatures from pseudobulk snRNA-seq data, with documented design decisions grounded in explicit inductive biases.
- A qualitative cell-type accuracy gradient consistent with the known psilocybin response hierarchy, reported alongside a direct quantitative test of HTR2A-gating that returns a result inconsistent with simple HTR2A-dose dependence.

- A novel finding that psilocybin-induced transcriptional downregulation is more stereotyped across individuals than upregulation, with a cortical depth gradient across excitatory subtypes.
- Characterization of drug-specific gene co-regulation modules recovered through attention analysis, including serotonin receptor and NMDA receptor pathway segregation.
- A formal hypergeometric test establishing that attention priorities show no statistically detectable overlap with classical DEG significance, with mechanistic discussion of what this does and does not imply.
- Honest documentation of architectural limitations — temporal structure failure and the attention-DEG divergence — with mechanistic explanations grounded in the training data design.

The delta expression encoder is designed as one module of a planned larger framework for psilocybin therapeutic response prediction, motivated by the observation that the causal path from drug administration to subjective experience passes through several distinct biological layers — pharmacokinetic exposure, receptor-level neuronal response, the resulting transcriptional change addressed here, and downstream biological and neural processes. This paper addresses the transcriptional layer in isolation; integration with the other layers is left to future work.

# Results

## 1. The delta expression encoder learns biologically coherent cell-type representations

The delta expression encoder was trained on 623 pseudobulk examples spanning 18 cell types, 2 drug conditions, and 6 timepoints. The final saved checkpoint (300 epochs) achieved 69.4% weighted accuracy (loss 0.7655); an earlier checkpoint at epoch 150 (70.3% accuracy, loss 0.7804) was used for the UMAP visualization; PCA-based silhouette scores were computed from an epoch-50 checkpoint; both are reported transparently.

Two baselines provide interpretive context for this metric: a majority-class classifier predicting NEUTRAL for all positions achieves 32.7% weighted accuracy, and a stratified random classifier achieves 32.8%, both reflecting the severe class imbalance in the data (94.0% NEUTRAL, 4.1% UP, 1.9% DOWN tokens among evaluable positions). The model's 69.4% represents a 2.1× improvement over either baseline.

Embedding separability was assessed in two complementary ways. Pooled across all 18 cell types, drug identity (psilocybin vs. ketamine) is linearly separable in the top 10 principal components of the 128-dimensional embedding space (silhouette = 0.1796), with cell type similarly separable (silhouette = 0.2081). In a lower-dimensional 2D UMAP projection, drug separability is weaker (silhouette = 0.0618) and cell-type separability is near zero (−0.0437); this is expected, since UMAP optimizes for 2D visualization rather than for preserving the full separability present in higher dimensions, and is reported for transparency rather than presented as the primary separability metric.

To directly assess what information is explicitly encoded as linearly separable structure in the embeddings, logistic regression probes were trained on frozen 128-dimensional embeddings using 5-fold cross-validation (balanced accuracy). Drug identity (psilocybin vs. ketamine) was perfectly decodable ($1.000 \pm 0.000$, n=2 classes). Cell type identity was equally perfect ($1.000 \pm 0.000$, n=18 classes). Timepoint was near-perfectly decodable ($0.974 \pm 0.016$, n=5 drug timepoints). Animal identity was at chance ($0.100 \pm 0.038$; theoretical chance for 34 animals $\approx 0.059$). The model has encoded drug context, cell-type identity, and temporal information in the embedding while discarding individual animal identity — precisely the correct inductive bias for a representation intended to generalize across subjects. These results confirm that the PCA silhouette-based separability reflects genuine linear structure, not a dimensionality-reduction artifact.

Hierarchical clustering of cell-type mean embeddings in PCA space (Ward linkage) revealed a dendrogram structure that closely mirrors known neurobiological taxonomy without this taxonomy being provided as a training signal. All eight excitatory cortical subtypes formed a single coherent cluster, cleanly separated from inhibitory and non-neuronal types — independently corroborating the biological class probe result above. Inhibitory interneurons and non-neuronal glial types intermixed within a second macro-cluster, consistent with the near-uniform attention entropy observed for both groups. The two highest-accuracy cell types — Sncg interneurons (99.0%) and endothelial cells (99.6%) — were jointly isolated at maximum distance from all other cell types, an emergent consequence of the training objective pushing trivially-classifiable cell types to the periphery of the representational space.

As a perturbation check, swapping drug labels on 20 population-mean examples and measuring the embedding shift yielded a mean L2 displacement of 14.16 ± 3.61 — comparable to within-drug variation (16.19) but substantially less than the natural cross-drug distance (21.65), confirming that drug identity is strongly encoded but not the sole organizing principle of the representation.

The embedding space is geometrically compact despite its 128-dimensional nominal size: 90% of variance is explained by 10 principal components, corresponding to an effective rank of approximately 10/128 — indicating the model has learned a low-dimensional manifold organized around the pharmacological and cell-type structure of the data.

As a test of whether the embeddings encode biological structure beyond the conditioning inputs, a probe for neuronal class (excitatory / inhibitory / non-neuronal — a three-class grouping never provided as a training label) achieved 1.000 ± 0.000 balanced accuracy (chance: 33.3%), confirming the model inferred biologically meaningful cell-class structure from expression patterns alone.

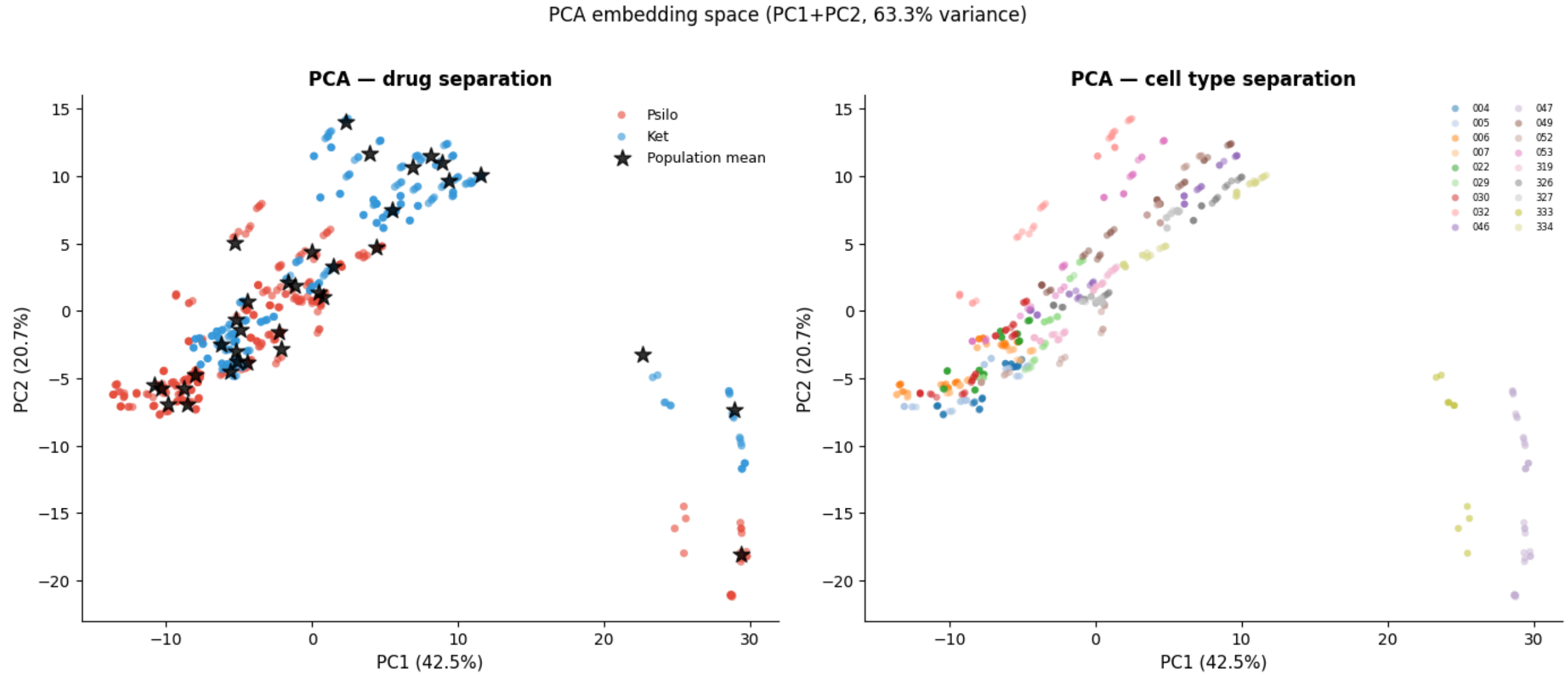

**Figure 1A.** *PCA of 128-dimensional embeddings (top 10 PCs, explaining 90% of variance). Left panel: drug separation. Right panel: cell-type separation across 18 cell types.*

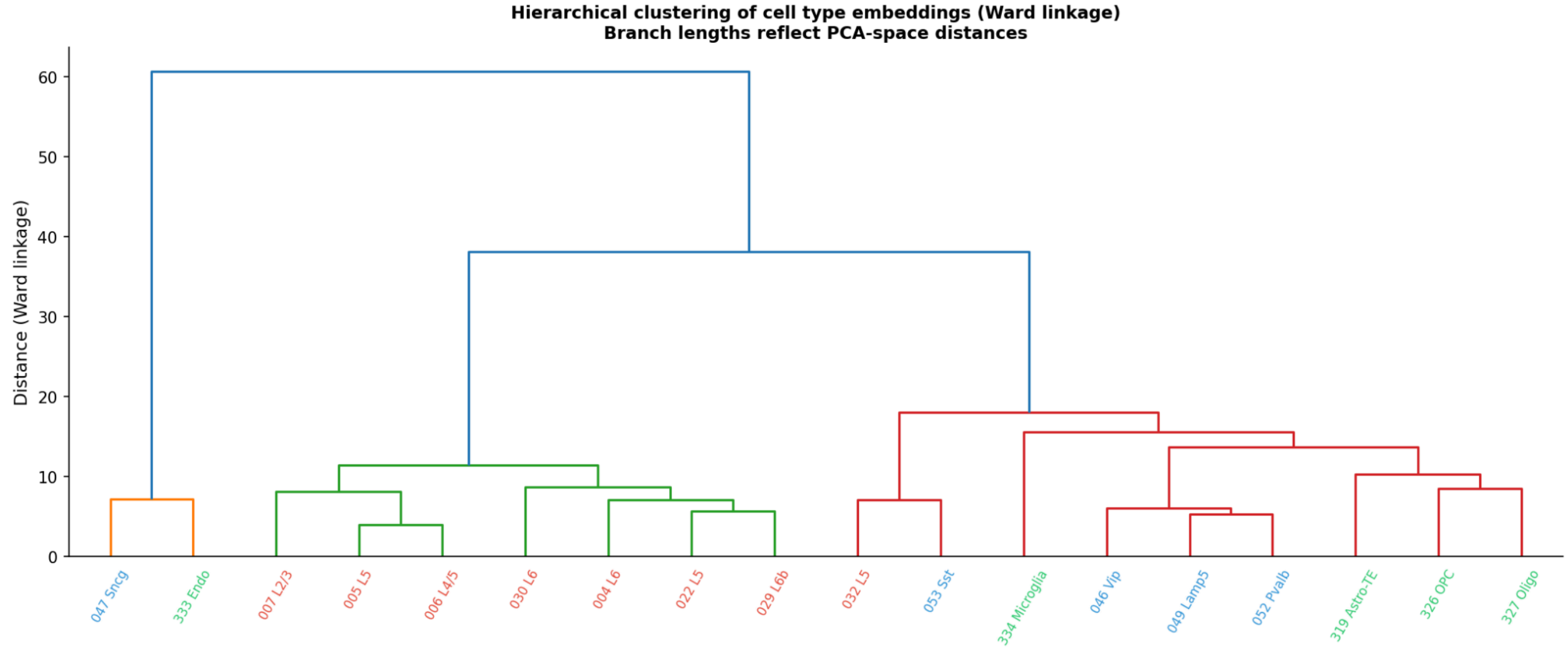

**Figure 1B.** *Hierarchical clustering of 18 cell-type mean embeddings in PCA space (Ward linkage, branch lengths reflect PCA distances). Excitatory cortical subtypes (green) form a single coherent cluster. Inhibitory interneurons and non-neuronal glial types intermix in a second macro-cluster (red). Sncg interneurons and endothelial cells (the two highest-accuracy cell types at 99.0% and 99.6%) are jointly isolated at maximum distance from all other types (blue), an emergent consequence of the training objective.*

## 2. Psilocybin-induced downregulation is more stereotyped than upregulation across individuals

The model's 2.2× gap in recall between DOWN-labeled (52.1%) and UP-labeled (23.7%) tokens has a straightforward mechanistic account: a model can only learn what is consistent across training examples. If psilocybin-induced upregulation is more variable across individuals than downregulation, the model will learn repressive programs more reliably than activating ones. Measuring this directly: genes labeled DOWN showed substantially lower inter-individual variance (mean std 0.1690, median 0.1443) than genes labeled UP (mean std 0.2403, median 0.2164), a significant difference confirmed by Mann-Whitney U test ($U=18615.0$, $p<0.0001$). The model's accuracy gap is not a modeling failure — it is a faithful reflection of the underlying biological asymmetry in transcriptional stereotypy.

This pattern held within each of eight excitatory cortical subtypes examined individually (L2/3 IT, L4/5 IT, two L5 subtypes, two L6 subtypes, L6b, and L5 NP), with the ratio of UP-variance to DOWN-variance ranging from approximately 1.07× to 1.61× across subtypes. Six of eight reached subtype-level significance ($p<0.05$) and the remaining two trended in the same direction.

One plausible biological account is that repressive transcriptional programs are often executed by constitutively active repressor complexes that respond uniformly to receptor stimulation, while activating programs may depend more on individual differences in chromatin accessibility, co-activator availability, or prior activity state. This account is biologically plausible and consistent with general principles of transcriptional regulation, but it has not been directly tested against chromatin or co-activator data here and should be read as a hypothesis motivated by this result, not a finding it establishes. To current knowledge, the underlying empirical observation — differential stereotypy of repressive versus activating transcriptional responses to psilocybin — has not been previously reported.

The structure of the model's classification errors provides independent, model-level evidence consistent with this directional asymmetry (Figure 2A). In the row-normalized confusion matrix computed across all 299,880 evaluable individual-example tokens, DOWN-labeled tokens achieve 52.1% recall while UP-labeled tokens achieve only 23.7% — a 2.2× gap consistent with DOWN being more stereotyped and therefore learnable from the training distribution. The error patterns differ directionally in a diagnostically meaningful way: when the model misclassifies a true DOWN token, it calls it NEUTRAL (33.4%) 2.3× more often than UP (14.5%), indicating conservative errors in the repressive direction. When the model misclassifies a true UP token, it calls it DOWN (33.0%) almost as often as NEUTRAL (43.3%), indicating that errors on activation are not directionally conservative and frequently invert the predicted sign. The model's error geometry mirrors the biological asymmetry characterized above: stereotyped repressive programs are more learnable and fail conservatively, while variable activating programs are less learnable and fail in both directions.

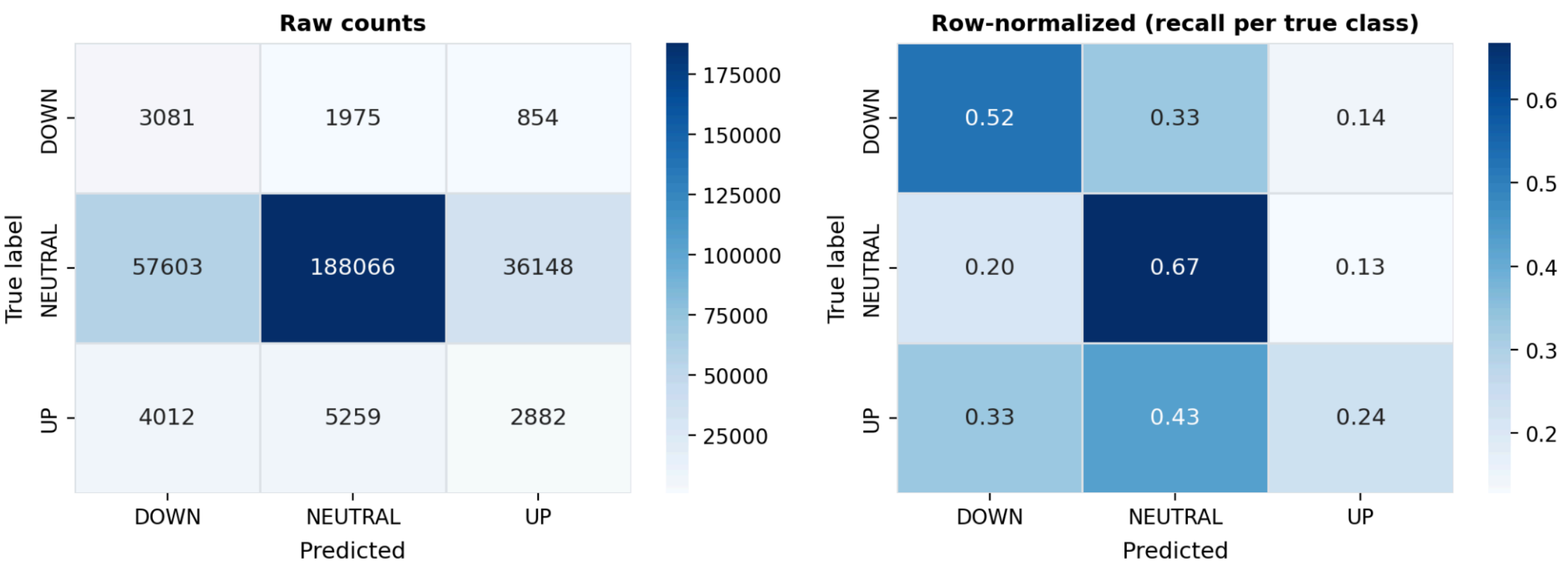

**Figure 2A.** *Classification confusion matrix. Left: raw token counts (DOWN/NEUTRAL/UP). Right: row-normalized recall per true class. DOWN recall (52.1%) substantially exceeds UP recall (23.7%); DOWN errors are directionally conservative (→ NEUTRAL) while UP errors are directionally incoherent (→ DOWN as frequently as NEUTRAL).*

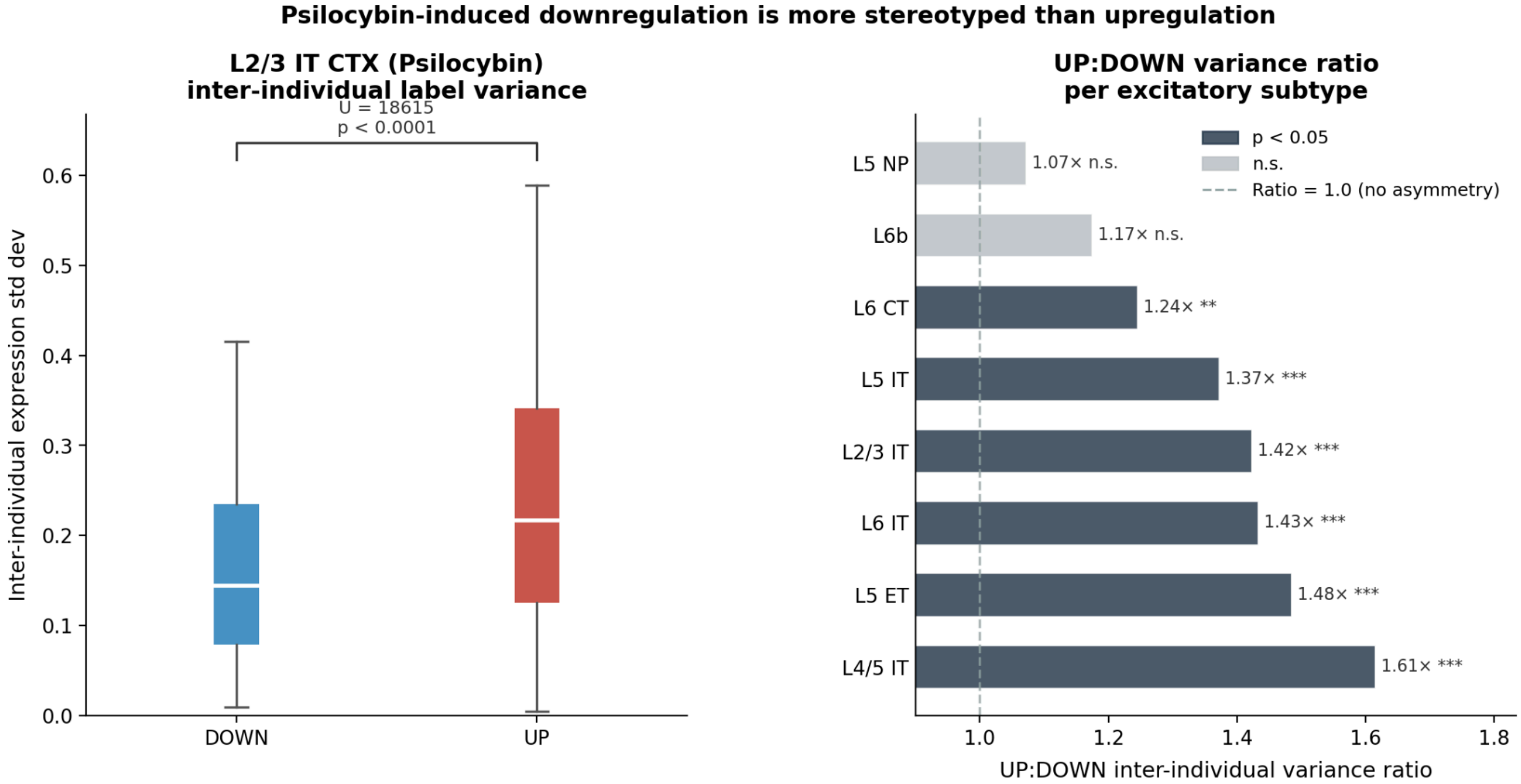

**Figure 2B.** *Distribution of inter-individual label variance for DOWN-regulated vs UP-regulated genes. Mann-Whitney U=18615.0, p<0.0001. Right panel: UP:DOWN variance ratio by excitatory subtype.*

### 3. Cell-type accuracy gradient is qualitatively consistent with known biology; a direct quantitative test of HTR2A-gating is not

Per-cell-type weighted classification accuracy ranged from 28.3% (L2/3 IT cortical glutamatergic neurons) to 99.6% (endothelial cells). The three lowest-accuracy cell types — L2/3 IT (28.3%), L4/5 IT (32.4%), and L6 IT (34.7%) — are excitatory intratelencephalic neurons, among the primary HTR2A-expressing psilocybin targets. The three highest-accuracy cell types — endothelial (99.6%), Sncg interneurons (99.0%), microglia (98.6%) — are non-neuronal or a molecularly distinct interneuron population. Attention entropy analysis revealed that the model applies near-uniform attention for both non-neuronal (mean entropy 5.91 ± 0.15, 94.7% of theoretical maximum) and excitatory (5.77 ± 0.23, 92.5% of maximum) examples, with non-neuronal entropy significantly higher (Mann-Whitney U=1074.0, p=0.0085). The near-ceiling entropy for non-neuronal types indicates the model's high accuracy on these cell types reflects a learned heuristic — near-universal NEUTRAL prediction — rather than gene-level selective inference. This does not affect the excitatory cell type analyses, where the pharmacologically relevant biology is concentrated.

This ordering is qualitatively consistent with cell types undergoing more psilocybin-induced transcriptional change being harder for the model to classify, and is now directly confirmed by a Spearman correlation against independent per-cell-type DEG counts from the Liao 2025 dataset: model accuracy showed a significant negative relationship with the number of significant psilocybin DEGs ($\rho = -0.885$, $p < 0.001$; Figure 3A), providing quantitative support for the hypothesis that high transcriptional response magnitude corresponds to a harder classification problem.

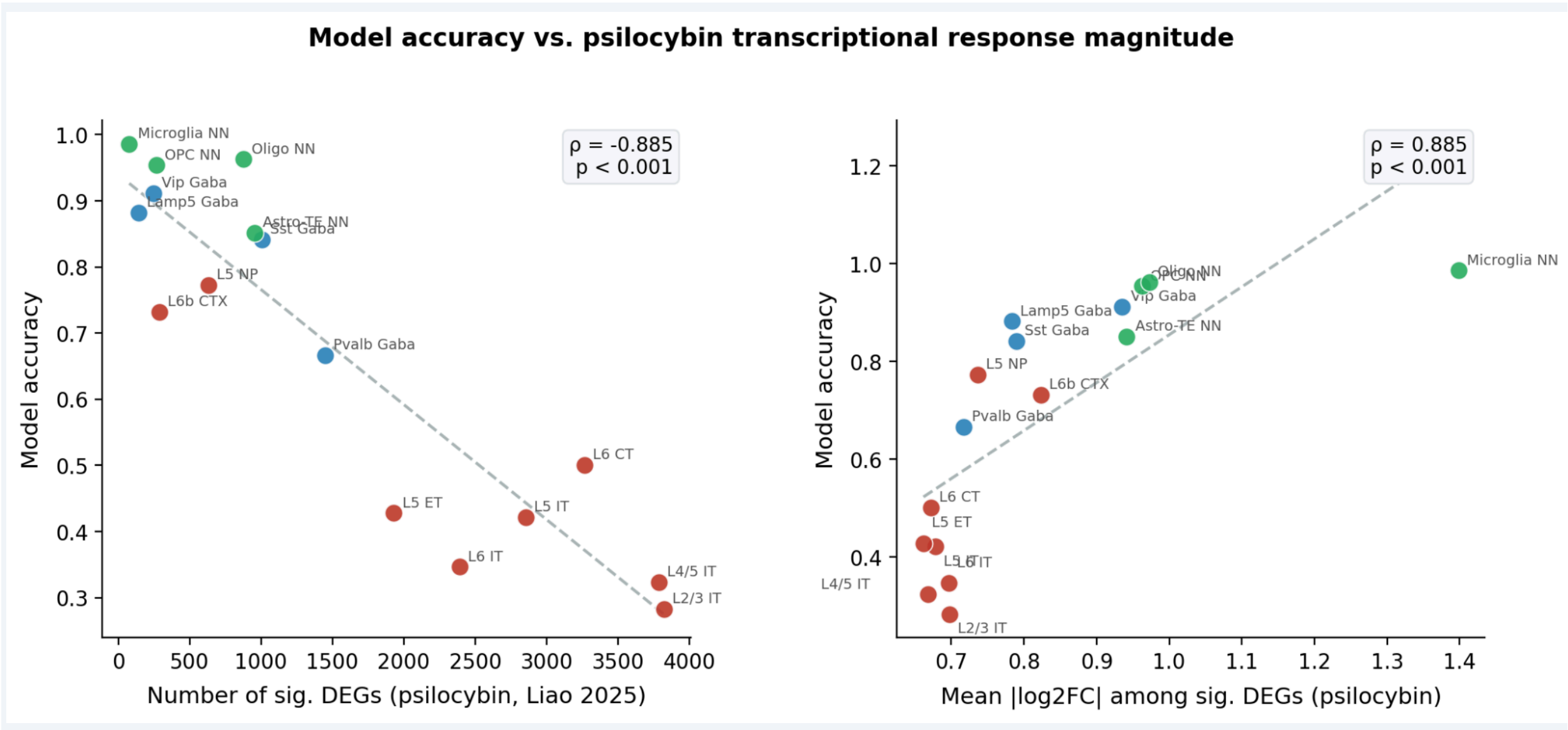

**Figure 3A.** *Quantifying the accuracy–response-magnitude relationship across cell types. Left: per-cell-type model accuracy versus number of significant psilocybin DEGs from Liao et al. 2025 (Spearman $\rho = -0.885$, $p < 0.001$). Right: model accuracy versus mean absolute log2FC among significant DEGs (Spearman $\rho = +0.885$, $p < 0.001$; the positive correlation reflects non-neuronal cell types having both large per-gene effects and near-perfect classification accuracy, while excitatory neurons combine high DEG count with lower per-gene magnitude). Color: red = excitatory (Glut), blue = inhibitory (Gaba), green = non-neuronal.*

Shao et al. 2025 demonstrated experimentally that psilocybin's lasting behavioral effects require intact HTR2A receptors specifically in pyramidal cell types. The HTR2A-gating hypothesis, in its simplest form, predicts that the strength of psilocybin's effect should scale with how much HTR2A a cell type expresses: more receptor → stronger activation → larger transcriptional response → more drug-specific molecular signature. If this holds, cell types with higher baseline HTR2A expression should show a more psilocybin-distinctive pattern in the model's embedding space — operationalized here as the drug-separation silhouette score (how far apart psilocybin vs. ketamine embeddings are for that cell type in PCA space, pooled across animals). This is the right operationalization because a larger, more psilocybin-specific transcriptional signature should correspond to embeddings that cluster more tightly by drug identity for that cell type. To test this directly, I correlated each cell type's baseline (drug-naïve) HTR2A expression against its drug-separation silhouette score. This returned a **significant negative** correlation (Spearman $\rho = -0.7088$, $p = 0.0021$, n=16 cell types): cell types with *higher* baseline HTR2A expression showed *weaker*, not stronger, drug-response separability in the model's embedding space. This is the opposite direction predicted by a simple model in which higher HTR2A expression should produce a larger, more separable psilocybin-specific transcriptional signature.

A second, complementary test correlated baseline HTR2A expression directly against measured psilocybin response magnitude in the underlying DEG data (independent of the model). HTR2A expression showed no significant relationship with DEG count ($\rho = 0.3706$, $p = 0.1577$) and a significant ***negative*** relationship with mean absolute log fold-change among significant DEGs ($\rho = -0.6559$, $p = 0.0058$) — the only one of the two correlations to reach significance, and in the negative direction. Neither result supports a simple monotonic HTR2A-dose-dependent model of psilocybin's transcriptional effect. A supplementary test using integrated baseline expression across all four serotonin receptor genes present in the panel (Htr1f, Htr2a, Htr1a, Htr2c) also showed no significant relationship with model accuracy ($\rho = -0.236$, $p = 0.484$), confirming that the null result is not specific to Htr2a alone.

A further test examined whether baseline HTR2A expression predicts drug-axis projection at the level of individual animals rather than cell-type averages. No consistent positive signal was found across excitatory subtypes (n_pos = 2/11 cell types, n_neg = 8/11), confirming that the absence of HTR2A-gating support is not an artifact of aggregation to the cell-type level.

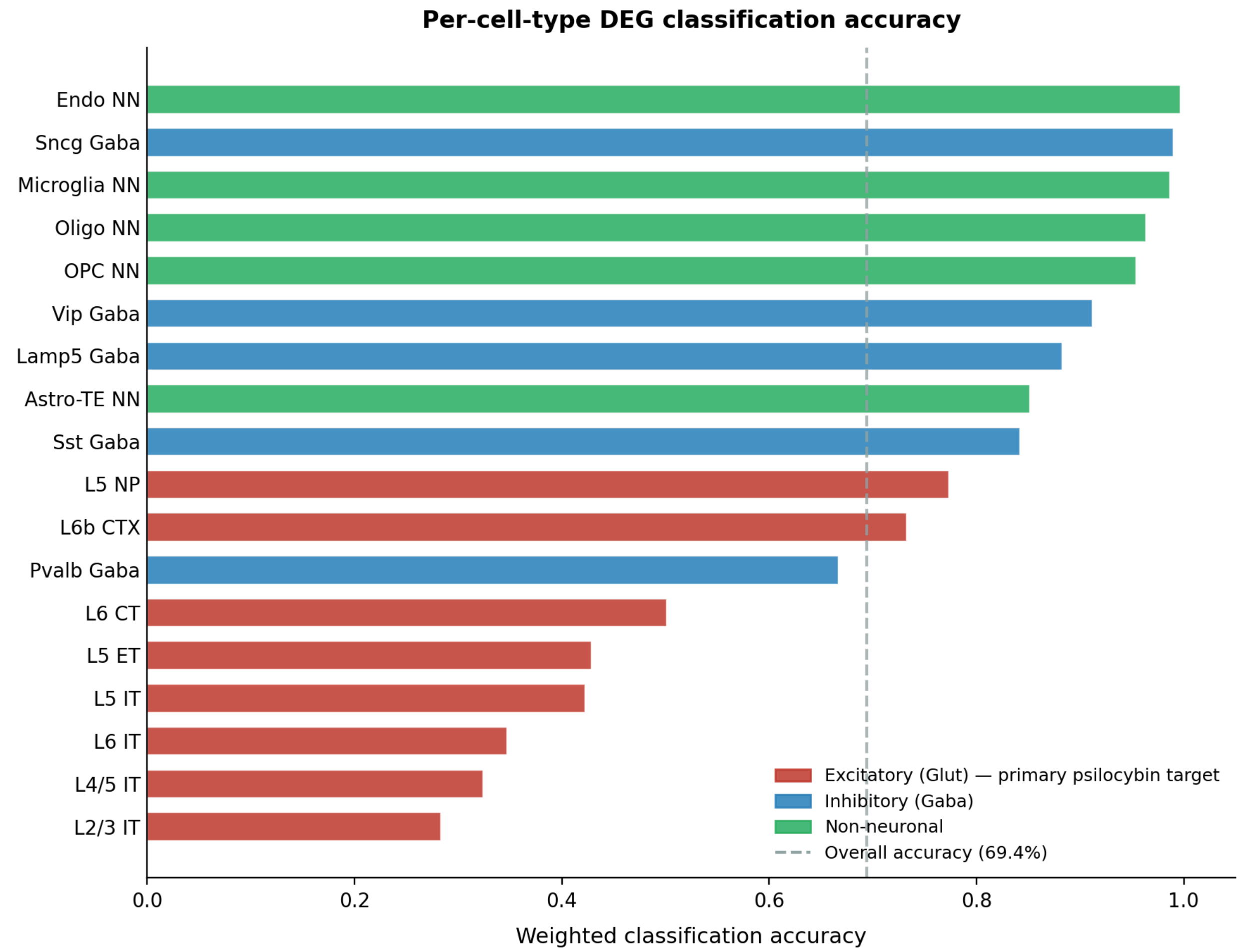

**Figure 3B.** *Per-cell-type weighted accuracy, sorted from lowest to highest, color-coded by cell class.*

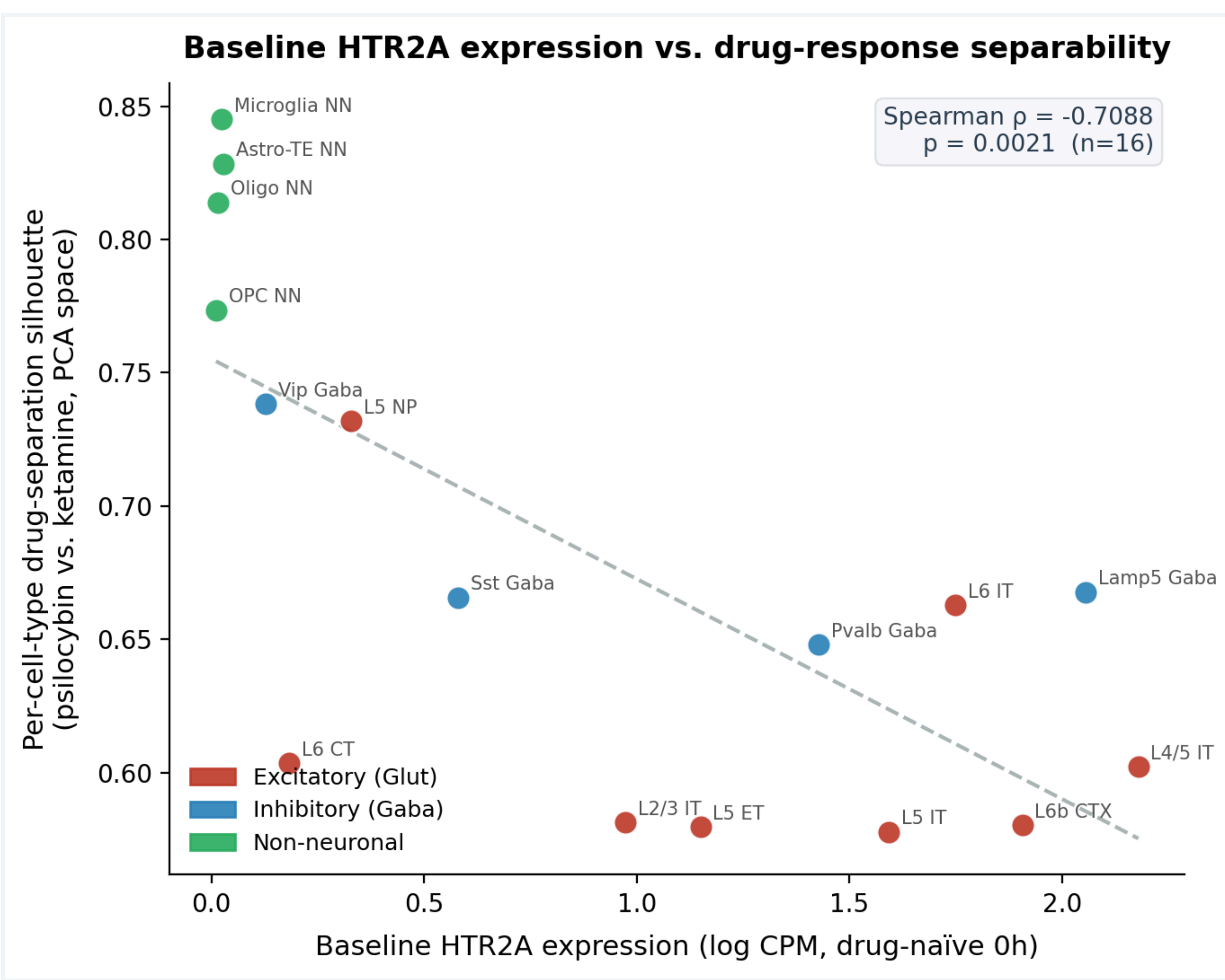

**Figure 3C.** *Baseline HTR2A expression vs. drug-response silhouette score per cell type. Spearman ρ = −0.7088, p = 0.0021 (n=16 cell types).*

### 4. Attention analysis recovers drug-specific gene co-regulation modules

The model's gene-gene attention matrix — attention weight summed across layers, heads, and timepoints, then averaged across examples of a given cell type — was used to identify co-regulated gene modules. Hierarchical clustering (Ward linkage, 80th percentile sparsification) produced 15 clusters, of which 14 were annotatable to known biological pathways (GO/KEGG/Reactome, g:Profiler, Benjamini-Hochberg FDR $< 0.05$). Because the underlying 510-gene panel was itself selected for differential expression significance rather than drawn at random from the genome, a baseline annotation rate above what would be seen for randomly chosen genome-wide genes is expected; the informative test is whether the *specific* pathways recovered correspond to known, drug-specific pharmacology.

Recovered modules included serotonin receptor co-regulation, alkaloid response, calcium signaling (KEGG $p = 1.83 \times 10^{-4}$), postsynaptic membrane potential regulation, NTRK2/BDNF signaling, and axon extension / PI3K-Akt signaling. Psilocybin-dominant clusters (P/K attention ratio > 1.3×) included the serotonin receptor cluster (C2, 1.67×) and alkaloid response cluster (C8, 1.52×), matching psilocybin's serotonergic mechanism. Ketamine-dominant clusters (P/K ratio < 0.75×) included kainate receptor activation (C12, 0.57×) and cellular component maintenance / postsynaptic spine organization (C4, 0.68×), and regulation of nervous system processes (C3, 0.64×), consistent with ketamine's glutamatergic

/ structural plasticity mechanism. This drug-specific structure emerged from attention weights alone; the model was never given either drug's receptor mechanism.

Inspection of gene membership within drug-selective clusters reveals pharmacologically coherent content that the model recovered without supervision. The primary ketamine-dominant cluster (C12, 0.57×) is anchored by Grin2b — the GluN2B NMDA receptor subunit that is ketamine's direct molecular target — alongside synaptic vesicle and calcium signaling genes (Syt1, Ryr2, Adcy1, Lrrk2). The primary psilocybin-dominant cluster (C2, 1.67×) contains Grin2a (GluN2A, a distinct NMDA subunit with different synaptic roles), Ntrk3, Erbb4, and Maob, reflecting serotonergic and plasticity-adjacent co-regulation. A second psilocybin-dominant cluster (C8, 1.52×) contains Fos and Bdnf — the canonical immediate-early gene and the primary mediator of psilocybin-induced synaptic plasticity — alongside interneuron markers Gad1, Vip, and Lamp5. The model's separation of Grin2b (ketamine-dominant) from Grin2a (psilocybin-dominant) is particularly notable: these two NMDA receptor subunits differ in synaptic localization and developmental expression, and their segregation into drug-specific attention clusters emerged from expression co-variation alone.

These recoveries constitute prospective validations against established pharmacology: Grin2b's ketamine-dominance and Fos/Bdnf's psilocybin-dominance are ground truths in the neuropharmacology literature, recovered here from expression co-variation alone without any mechanistic supervision.

To test whether the drug-specific cluster structure holds within the primary psilocybin target cell type specifically, the co-attention analysis was repeated using L2/3 IT neuron examples exclusively (n=17 psilocybin, n=17 ketamine). The resulting cluster structure differed from the global analysis: the most psilocybin-dominant cluster within L2/3 IT (C7, ratio=1.858×) contained cytoskeletal and glial transporter genes (Slc1a2, Lama2, Kcnmb2) rather than canonical serotonin receptors, while the most ketamine-dominant cluster (C12, ratio=0.598×) contained synaptic vesicle and ion channel genes (Syt1, Cacnb4, Slc6a17). This indicates that the globally recovered serotonin receptor co-regulation module reflects signal averaged across multiple cell types rather than a feature specific to L2/3 IT neurons — a distinction that contextualizes rather than undermines the global finding, since L2/3 IT-specific transcriptional response to psilocybin may be mediated through downstream effectors rather than through receptor co-regulation itself.

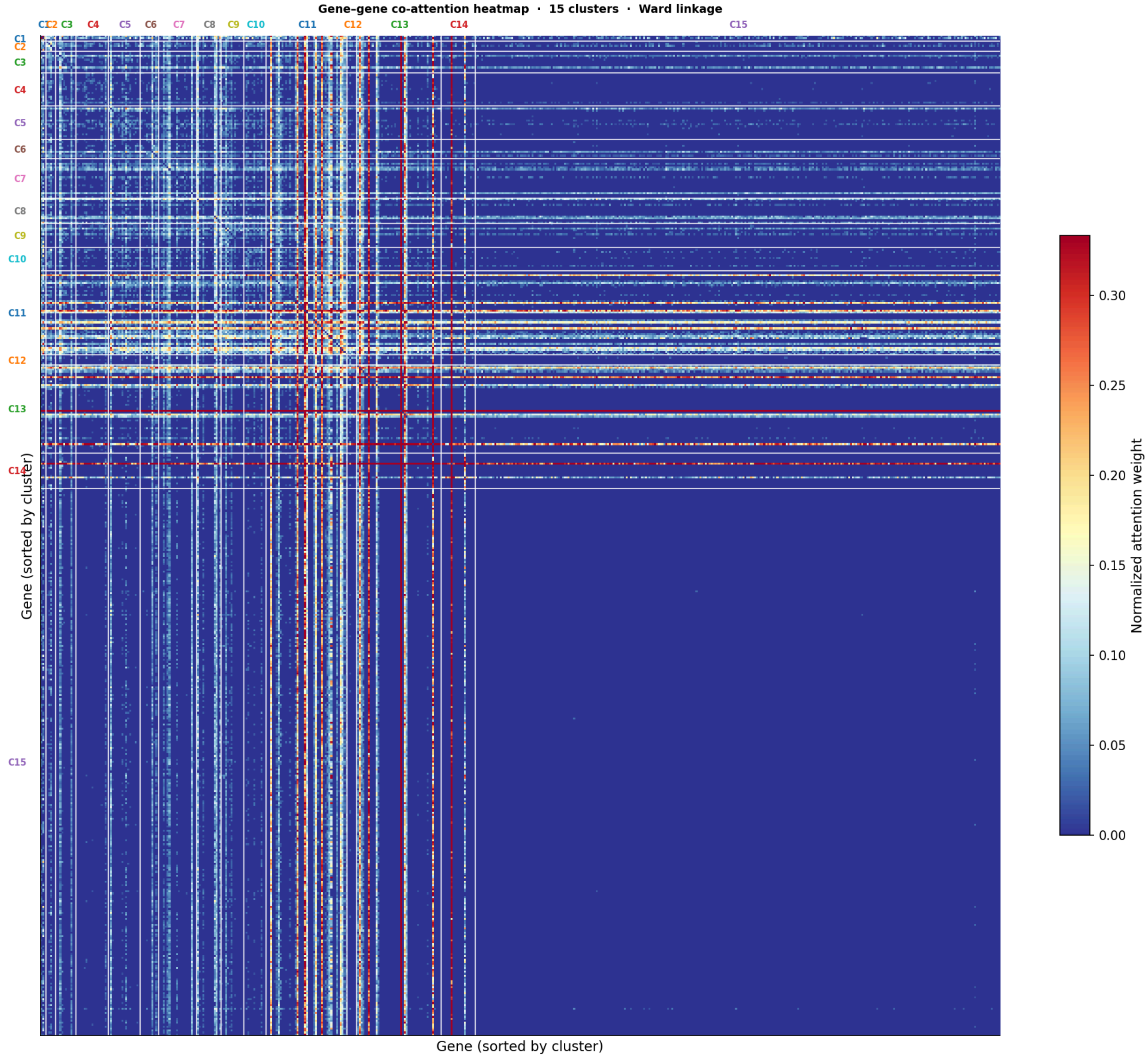

**Figure 4A.** *Gene-gene co-attention heatmap (510 × 510, Ward linkage, 80th percentile sparsified). Genes reordered by cluster membership. Cluster numbers C1–C15 labelled on axes; pathway annotations and P/K drug-selectivity ratios are reported in Results text, not overlaid on figure.*

The distribution of attention across the model's four layers was examined to understand how the representation builds up across depth. For a focal gene (Htr2a, the primary psilocybin receptor), incoming attention from the first 50 genes in the panel was tracked per layer (Figure 4B). Layers 1 and 2 show diffuse, near-uniform attention (magnitudes ~0.0003–0.0005 across most positions). Layer 3 exhibits a qualitative shift: magnitudes jump approximately 3-fold (reaching ~0.0016) while the distribution becomes sharply sparse, with most genes receiving near-zero attention and a small number of positions receiving concentrated weight. Layer 4 shows reduced overall magnitudes (~0.000175) but retains the sparse pattern from Layer 3. This two-phase profile — broad context assembly in Layers 1–2, sharp gene selection in Layer 3, and refinement in Layer 4 — indicates that drug-specific gene prioritization emerges primarily at Layer 3 rather than accumulating gradually, a processing structure consistent with interpretability findings in other deep Transformer models applied to biological sequence data.

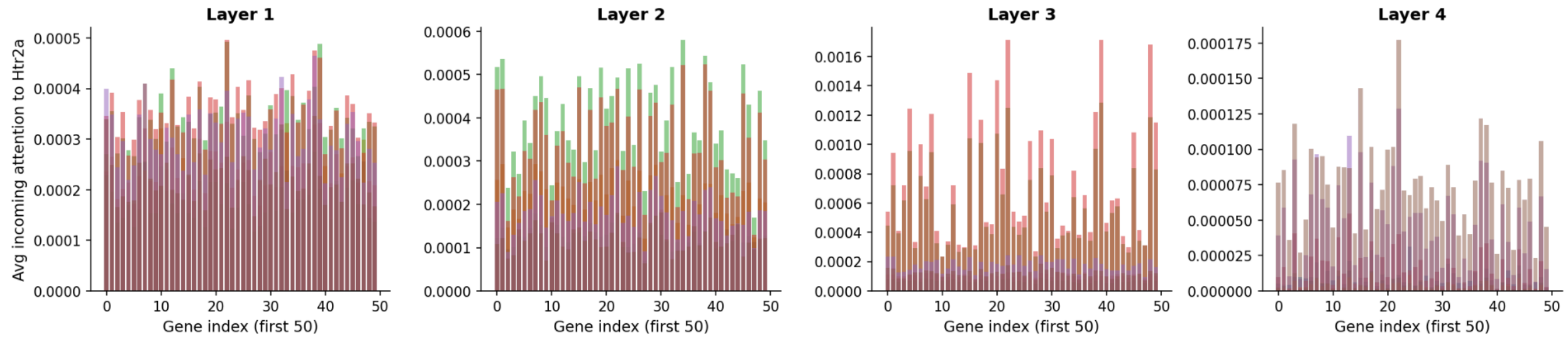

**Figure 4B.** *Layer-by-layer incoming attention to Htr2a across the 4 Transformer encoder layers (first 50 genes shown). Layers 1–2 show diffuse, near-uniform attention; Layer 3 shows sharp concentration with ~3× magnitude increase; Layer 4 refines.*

## 5. Attention priorities show no detectable enrichment for classical DEG significance

The top 30 genes by temporal attention coefficient of variation (genes whose attention weight varies most across the 6 timepoints) were compared against Liao et al.'s statistically significant DEGs in L2/3 IT neurons at 1h and 72h, using a hypergeometric enrichment test (universe = 510 genes). The observed overlap was **zero genes at both timepoints** (expected by chance: 1.3 genes at 1h, 0.7 genes at 72h; hypergeometric p = 1.0 at both timepoints — the overlap is not merely non-significant, it is below the chance expectation). Extending the comparison to the top 50 and top 100 attention genes did not change this conclusion.

This indicates the genes the model's attention prioritizes are not the genes with the largest classical fold-changes, at the level of statistical resolution this test can detect. Two readings of this result are possible. One is that the model has learned to track a real, different signal — genes that are informative for classification given full context — that is genuinely distinct from large-effect-size DEGs. The other is that the model has not learned anything pharmacologically meaningful in its temporal attention pattern specifically. The cell-type accuracy ordering (Results 3) and the drug-specific pathway segregation (Results 4) are both evidence the model's broader representation tracks real biology; this section's null result is reported as a direct, limiting fact about the temporal attention analysis specifically, not explained away by those other results, since they test different aspects of the model.

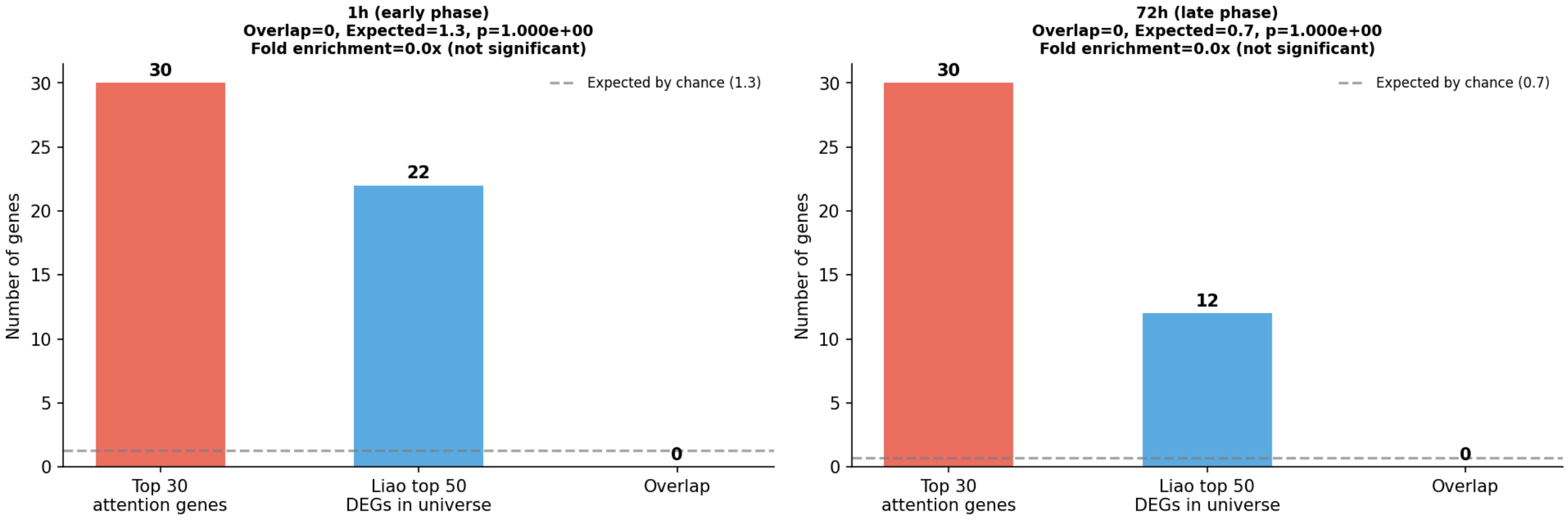

**Figure 5.** *Attention gene priorities show no overlap with classical DEG significance at 1h and 72h in L2/3 IT neurons. Top 30 genes ranked by temporal attention coefficient of variation (CV) compared against Liao et al.'s top statistically significant DEGs (by padj, then |log2FC|) at 1h and 72h. Observed overlap: 0 genes at both timepoints (expected by chance: 1.3 at 1h, 0.7 at 72h; hypergeometric p = 1.0 at both timepoints). The model's attention priorities track a signal distinct from classical differential expression significance.*

## 6. Attention flags a candidate sustained suppression pattern in Htr1f

During systematic temporal attention analysis (Results 5), one gene — Htr1f, the serotonin 1F receptor — exhibited an anomalously high attention coefficient of variation (CV = 1.668), ranking 4th of 510 genes. Examining its attention values explicitly: at 72h, Htr1f's attention weight is 0.007802, compared to 0.000241 at 1h and 0.000832 at 24h — a ~9× elevation above the 24h baseline and ~32× above 1h. No other serotonin receptor in the panel (Htr2a, Htr1a, Htr2c) shows a comparable temporal spike. Inspecting Htr1f's expression trajectory directly from the database: across all cell types, mean baseline expression is 2.692 (log CPM); under psilocybin it falls sharply to 2.025 at 1h, recovers partially to 2.618–2.542 at 2–4h, and declines a second time to 2.362 at 72h. Ketamine shows no comparable trajectory, remaining within a narrow 2.49–2.64 range across all timepoints. In L2/3 IT neurons specifically (baseline 0.938), the 1h suppression is an established, statistically significant DEG ($p<0.001$ across 7 cell types); the 72h dip does not reach significance in the standard per-timepoint DEG test at this cell type ($p=0.22$, likely underpowered at $n=4$ animals per timepoint), but psilocybin shows lower expression than ketamine at 72h in 14 of 16 cell types examined.

Further, Htr1f's outgoing attention at 72h is dominated by attention to Kcnq1ot1 at 2h (weight 0.0747), a long noncoding RNA involved in gene silencing through imprinting regulation, followed by self-attention (0.0427) and attention to Eml5 (0.0342), a microtubule-associated gene. The co-attention cluster analysis independently placed Htr1f in Cluster 2 (serotonin receptors, P/K ratio 1.665× — the most psilocybin-dominant cluster in the analysis), alongside Tek (a receptor tyrosine kinase expressed in vascular endothelium). This is convergent evidence from two independent analyses — temporal attention flagging and co-regulation clustering — pointing to the same gene.

This finding should be read as a low-confidence, hypothesis-generating result for two reasons disclosed directly by collaborators with access to complementary data. First, the Allen Institute's SMART-Seq reference dataset shows substantial baseline Htr1f expression in mouse frontal cortex, but the newer MERFISH reference does not — a platform-dependent discrepancy serious enough that Htr1f was deliberately excluded from the corresponding figure in Liao et al.'s own preprint, and one that means the apparent 72h suppression could in part reflect a sequencing-method artifact rather than true biology. Second, Htr1f's function in this circuit is not well characterized in the literature.

The convergence of two independent analytical routes — temporal attention flagging (CV = 1.668, ranking 4th of 510 genes) and co-attention cluster placement (Cluster 2, P/K ratio 1.665×) — on the same gene suggests this is a signal worth following up, even at low confidence.

**Figure 6.** *Htr1f expression trajectory under psilocybin vs. ketamine. Left panel: L2/3 IT neurons, with the 1h suppression (established DEG, p<0.001) and 72h dip (model-flagged, p=0.22 n.s.) annotated. Right panel: all-cell-type average, with the 72h effect directionally consistent across 14/16 cell types.*

## 7. Temporal structure is not recovered, and the mechanistic reason is the training data design

Liao et al. 2025 identified a biphasic temporal structure of psilocybin's transcriptional response: a strong early wave at 1–2 hours, near-silence at 4–24 hours, and a return at 72 hours. The model failed to recover this structure in population-mean signed-score analysis.

To quantify this failure, two direct tests were performed against published Liao et al. 2025 figures, using L2/3 IT neurons as the reference cell type. First, the Pearson correlation between per-gene psilocybin and ketamine signed model scores was computed at each timepoint. Liao et al. Fig 6c shows this correlation should be highest at the active timepoints (1h, 72h) and lowest at the quiet interval (4–24h); the model instead returns uniformly high correlation across all timepoints (r = 0.916, 0.909, 0.901, 0.897, 0.908 at 1h, 2h, 4h, 24h, and 72h respectively.). The peak-to-trough difference is 0.019 r units — directionally correct (1h and 72h are marginally higher than 24h) but negligible in magnitude, establishing a quantitative upper bound on the temporal structure the model has learned: essentially none. Second, the fraction of genes receiving a confident directional score (|score| > 0.1) was computed per timepoint. Liao et al. Fig 5b shows two directional peaks for psilocybin at 1–2h and 72h; the model instead shows a monotonically rising fraction from 1h to 24h followed by a drop — the opposite temporal shape. Both results are shown in Figure 7.

This is a direct, mechanistic consequence of the training data structure, not of model capacity. The Liao et al. 2025 dataset is between-subjects: an animal can only be sacrificed once, so each individual animal contributes a real measurement at exactly one post-administration timepoint. No individual training example therefore contains a real temporal trajectory. The only training signal carrying real across-timepoint information comes from 35 population-mean examples (one per drug × cell-type combination) — a small fraction of the 623 total training examples, and evidently insufficient to teach the biphasic shape.

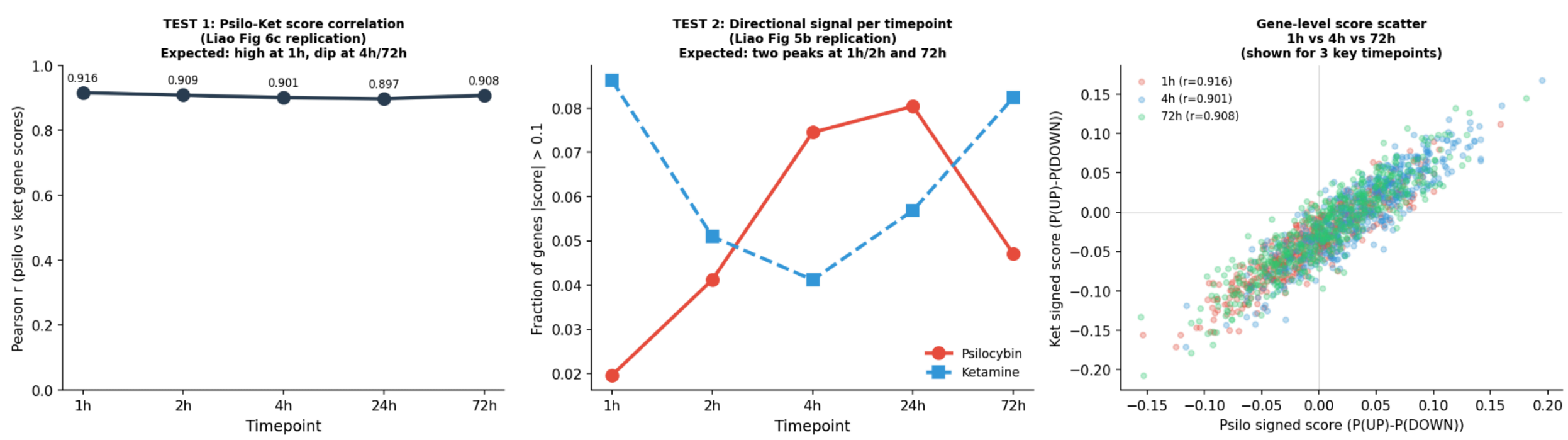

**Figure 7.** *Quantitative test of temporal structure recovery, L2/3 IT CTX Glut neurons. Left: Pearson correlation between per-gene psilocybin and ketamine signed model scores at each timepoint. Expected from Liao et al. 2025 Fig 6c: highest correlation at active timepoints (1h, 72h), lowest at quiet interval (4–24h). Model shows flat ~0.90 across all timepoints. Middle: Fraction of genes with |score| > 0.1 per timepoint. Expected from Liao et al. 2025 Fig 5b: two directional peaks for psilocybin at 1–2h and 72h. Model shows monotonically rising psilocybin signal through 24h then a drop — opposite temporal shape. Right: Gene-level scatter of psilocybin vs. ketamine signed scores at 1h, 4h, and 72h; all three timepoints cluster nearly identically (r ≈ 0.90), confirming the model does not temporally discriminate.*

# Methods

## Dataset

The training dataset is from Liao et al. 2025 *[Liao 2025]*, a single-nucleus RNA-seq study of mouse medial frontal cortex profiling psilocybin and ketamine administration across five post-administration timepoints (1h, 2h, 4h, 24h, 72h), plus drug-naïve control animals sacrificed at a single baseline (0h) timepoint. The updated AnnData file with Allen Institute cell-type annotations and animal ID metadata was used. Raw FASTQ data is available at SRA BioProject PRJNA1204073; processed data is available via Zenodo (DOI: 10.5281/zenodo.19666128).

## Gene panel

A panel of 510 genes was selected by combining statistical differential-expression criteria with a curated set of pharmacologically relevant candidate genes specified in advance — including HTR2A, HTR1A, HTR2C, TPH2, SLC6A4, COMT, MAOA, MAOB, BDNF, ARC, FOS, and ionotropic glutamate receptor subunits — to ensure pharmacologically central genes were retained regardless of statistical selection.

## Input representation

### *Pseudobulk construction*

For each combination of animal, cell type, drug condition, and timepoint, expression was averaged across all nuclei of that cell type from that animal at that timepoint, restricted to the 510-gene panel; combinations with fewer than 10 nuclei were excluded.

### *Training example construction*

Each training example corresponds to one animal × one cell type, represented as a 510 × 6 matrix (timepoints 0h through 72h). The 0h column for every example is the real population mean expression across all control animals for that cell type — control animals are only ever measured at 0h. The single drug timepoint column matching the animal's actual measured timepoint contains real individual pseudobulk expression. All remaining drug timepoint columns contain the sentinel value −1.0. Separately, 35 population-mean examples (one per drug × cell-type combination, with real population-mean values at every timepoint) provide the only training signal with complete temporal coverage. The full training set comprised 588 individual examples and 35 population-mean examples (623 total).

### *Sentinel value rationale*

Earlier model iterations imputed unmeasured timepoints with population means rather than a sentinel value, which allowed the model to partially infer cell type from the imputed values themselves rather than learning from the genuinely measured timepoint. The sentinel value (−1.0, never produced by real data) removes this shortcut.

To validate that timepoint information is actively used rather than ignored, sentinel injection analysis confirmed that ablating each drug timepoint column shifts the mean-pooled embedding by 0.68–1.91 units (L2/3 IT Psilo example; 2h most critical at 1.91, 72h least at 0.68), confirming the model extracts signal from whichever real timepoint is present.

### *Per-example normalization*

Each 510 × 6 input matrix was normalized to zero mean and unit standard deviation across non-sentinel values prior to projection, removing between-example magnitude variance that could otherwise let the model infer cell type from overall expression magnitude.

## Model architecture

The DeltaExpressionEncoder is a 4-layer Transformer encoder operating on a flattened sequence of 3,060 tokens (510 genes × 6 timepoints). Each token is the sum of: (1) the scalar expression value, projected to $d^{model}$ = 128 dimensions; (2) a learned gene identity embedding (Embedding[510, 128]); (3) a fixed sinusoidal positional encoding over the 6 timepoints; and (4) cell type and drug conditioning embeddings (Embedding[18, 128] and Embedding[2, 128]), broadcast to every token. Sentinel positions are replaced with a learned mask-token vector before projection.

The encoder has 4 layers, 4 attention heads, feedforward dimension 512, dropout 0.05, and pre-layer normalization, with residual connections around both sub-layers in each block. Token dimensionality (128) is fixed across all layers, a requirement of the residual connections. Within each layer, self-attention lets every token incorporate information from every other token; the feedforward sub-layer then transforms each token independently. A linear classification head maps each token's final representation to 4 logits (DOWN/NEUTRAL/UP/BASELINE). For representation analyses, a single embedding per example is obtained by mean-pooling across all 3,060 final-layer token representations. Total model parameters: 862,084.

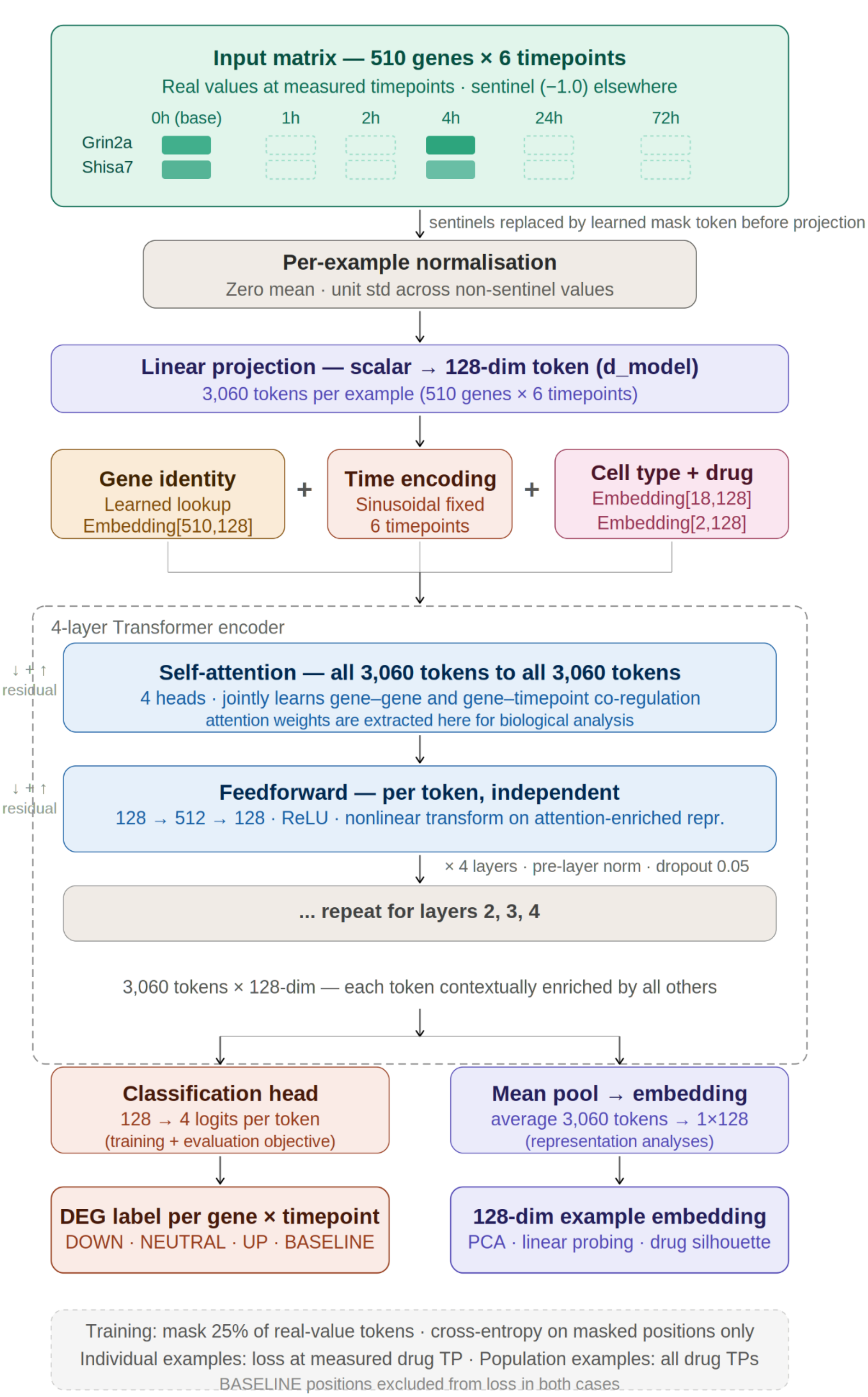

**Figure 8.** *Model architecture schematic: input construction, token assembly, 4-layer Transformer encoder, classification and embedding outputs.*

## Training

At each training step, 25% of real-value tokens were randomly masked; sentinel positions were never masked. Cross-entropy loss with inverse-frequency class weighting (DOWN: 2.52, UP: 1.19, NEUTRAL: 0.05, BASELINE: 0.24) was applied only to masked tokens, restricted to the animal's one real measured drug timepoint for individual examples, or all drug timepoints for population-mean examples; the 0h/BASELINE column is excluded from the loss in both cases, since its role is to serve as a known reference value rather than a quantity to be inferred.

Given the severe class imbalance (94.0% of evaluable tokens are NEUTRAL), standard accuracy would be dominated by NEUTRAL prediction; weighted accuracy down-weights NEUTRAL (weight 0.05) and up-weights the pharmacologically informative classes (DOWN: 2.52, UP: 1.19), making it the appropriate primary metric.

AdamW optimizer (learning rate $1\times10^{-3}$, weight decay 0.01), linear warmup over 20 epochs, cosine decay to $1\times10^{-5}$ over 300 total epochs, trained on an NVIDIA L4 GPU. The final saved checkpoint (300 epochs) achieved 69.4% weighted accuracy (loss 0.7655); an earlier checkpoint at epoch 150 (70.3% accuracy, loss 0.7804) was used for UMAP and embedding analyses; epoch 50 was used for PCA-based silhouette scores, as disclosed in Results.

## Evaluation and analysis

### *Embedding probing*

PCA was applied to 128-dimensional mean-pooled embeddings; silhouette scores for drug and cell-type separability were computed both in the top-10-PC PCA subspace (capturing 90% of variance) and in a 2D UMAP projection, with both reported for transparency. Linear probes (logistic regression, C=1.0, max_iter=1000) were trained on frozen individual-example embeddings using 5-fold stratified cross-validation, with balanced accuracy as the metric, for four prediction targets: drug identity, cell type, timepoint, and animal identity. Drug swap perturbation was measured as the L2 distance between full-information and drug-swapped embeddings on 20 population-mean examples.

A biological class probe (excitatory / inhibitory / non-neuronal, never provided as a training label) was also trained using the same 5-fold CV procedure to test whether embeddings encode structure beyond the explicit conditioning inputs.

### *Attention analysis*

Attention weight matrices were extracted from all 4 layers (averaged across heads). For temporal analysis, attention received by each gene was averaged per timepoint and ranked by coefficient of variation across timepoints. For co-attention clustering, a 510×510 gene-gene matrix (summed across timepoints, layers, and heads, averaged across same-cell-type examples) was z-scored, sparsified at the 80th percentile, and clustered (Ward linkage). Pathway enrichment used g:Profiler (GO/KEGG/Reactome, Benjamini-Hochberg FDR<0.05).

### *HTR2A-gating tests*

Baseline (drug-naïve, 0h) HTR2A expression per cell type was correlated (Spearman) against: (1) per-cell-type drug-response silhouette score in PCA embedding space; (2) per-cell-type psilocybin DEG count; and (3) per-cell-type mean absolute log2 fold-change among significant psilocybin DEGs.

### *Attention-DEG enrichment test*

The top 30/50/100 genes by temporal attention coefficient of variation were tested for enrichment against Liao et al.'s top 50 significant DEGs (by padj, then |log2FC|) at 1h and 72h in L2/3 IT neurons, using a one-tailed hypergeometric test (universe = 510 genes).

***Variance asymmetry analysis***

For each gene labeled DOWN or UP in a given cell type, the standard deviation of that label across individual pseudobulk profiles was computed; a Mann-Whitney U test compared DOWN-label-variance and UP-label-variance distributions, both pooled and per excitatory subtype.

## Discussion

This work presents a Transformer-based delta expression encoder for psilocybin transcriptional response and reports its validation honestly, including a result that runs counter to a published hypothesis. Without pathway supervision, the model learns drug- and cell-type-separable representations, and recovers pharmacologically correct drug-specific gene modules through attention alone.

The HTR2A-gating result (Results 3) deserves direct discussion rather than omission. A simple account in which higher HTR2A expression produces a proportionally larger, more separable psilocybin response would predict a *positive* correlation between baseline HTR2A and both drug-response silhouette and DEG magnitude. Three independent tests instead found null or significantly *negative* relationships. This does not contradict Shao et al. 2025's causal finding that HTR2A and pyramidal cell types are necessary for psilocybin's behavioral effects — necessity does not imply a monotonic dose-response relationship at the transcriptional level, and cell types with the highest baseline HTR2A may already be near a response ceiling, or HTR2A's role may be permissive rather than rate-limiting for the transcriptional cascade measured here. But the data do not support treating HTR2A expression level as a simple linear predictor of transcriptional response magnitude, and the cell-type accuracy gradient (low accuracy in L2/3 IT, high in non-neuronal types) is now quantitatively confirmed against independent DEG counts (Spearman $\rho = -0.885$, $p < 0.001$; Figure 3A), though the HTR2A-specific gating hypothesis remains unsupported by the data.

Two additional mechanisms could produce the observed negative correlation specifically rather than a null result. First, high-HTR2A pyramidal neurons co-express the highest levels of NMDA receptor subunits in cortex, meaning both psilocybin and ketamine produce large transcriptional responses in the same cells — geometrically compressing drug-separation in embedding space even when the underlying mechanisms differ. Second, these cell types may operate near a transcriptional response ceiling at the doses studied, producing large but convergent perturbation signatures regardless of drug identity; lower-HTR2A cell types, operating further from ceiling, may produce more drug-discriminative responses simply because they have more dynamic range available.

The DOWN/UP variance asymmetry (Results 2) is a novel, independently verifiable empirical observation, robust across excitatory subtypes. The repressor-complex mechanistic account offered for it is plausible but untested directly here and should be treated as a hypothesis the result motivates, not a finding it establishes.

The recovery of drug-specific, pharmacologically correct gene co-regulation modules through attention (Results 4) is, on balance, the strongest evidence that the model's representation captures real pharmacological structure, since the model was never given either drug's receptor mechanism. This sits alongside the null result in Results 5: attention priorities do not detectably overlap with classical DEG significance. These two results are not in tension — a gene can be important to the model's learned co-regulation structure without having the single largest fold-change — but the null enrichment result is reported as a direct limiting fact about what temporal attention analysis alone can support, rather than reasoned away.

A deeper translational challenge concerns species generalization. This model is trained on mouse mPFC, and direct application to human therapeutic response prediction requires bridging a gap that human psilocybin transcriptomics data cannot yet close — no dataset of the required scale exists. One path forward is indirect: rather than requiring human transcriptomics, a multi-level computational framework could ground mouse-derived molecular representations in human biological signals that are measurable — neural dynamics, pharmacokinetic profiles, or genomic variation in receptor targets — using the pharmacological perturbation itself as the cross-level anchor. Psilocybin's serotonergic mechanism operates through conserved receptor targets across species, providing a biological basis for alignment that does not depend on matched transcriptomics data. A dedicated cross-species translation module — trained to map mouse cell-type embeddings to human cell-type expression profiles using orthologous gene expression as the alignment signal, drawing on resources like the Allen Brain Cell Atlas and Human Cell Atlas — represents a natural extension of this framework and could be reused across any experimental model built on rodent transcriptomics data.

### Limitations and future directions

The failure to recover psilocybin's biphasic temporal structure is a direct consequence of the between-subjects training data design: no individual animal can be measured at more than one timepoint, so no training example contains a real multi-timepoint trajectory. Addressing this would require within-subject longitudinal data, which does not currently exist for psilocybin in mouse.

The near-uniform attention observed for non-neuronal cell types (94.7% of maximum entropy) suggests the model relies on a heuristic rather than gene-level inference for these cell types, and their high classification accuracy should be interpreted accordingly.

The HTR2A-gating and attention-DEG enrichment results in this paper are negative or null with respect to specific hypotheses; they constrain rather than support certain interpretations of what the model has learned, and are reported as such rather than reframed as confirmatory.

The model is trained on mouse data only. Extension to human therapeutic response prediction would require a human ortholog mapping analysis and ideally human psilocybin transcriptomics data, which does not yet exist at the scale required.

Future work will extend this encoder toward a broader computational framework for modeling individual differences in psilocybin therapeutic response, integrating additional biological modalities not addressed here. Cross-dataset validation on independent psilocybin transcriptomics datasets is also planned.

## Acknowledgements

I thank Dr. Alex Kwan (Kwan Lab, Cornell University / University of Michigan) and Ethan O'Farrell for providing the updated psilo-seq AnnData with animal ID metadata, for responsive engagement with biological questions arising during model development, for the technical caveat regarding the Htr1f SMART-Seq/MERFISH discrepancy that appropriately constrained the interpretation of that finding, and for reviewing a draft of this manuscript.

The author used Claude (Anthropic) as an AI writing assistant for manuscript drafting, editing, and revision. The author takes full responsibility for the accuracy, integrity, and originality of all content. The author declares no competing interests.

## Data and Code Availability

The Liao et al. 2025 psilo-seq dataset is available via Zenodo (raw FASTQ at SRA BioProject PRJNA1204073; processed AnnData DOI at 10.5281/zenodo.19666128). Model code, training scripts, and evaluation notebooks are publicly available at GitHub (*https://github.com/saijaya/delta-expression-encoder*). Model checkpoints are deposited at Zenodo (DOI: 10.5281/zenodo.21095034).

## Figure List

Figure 7. Temporal grammar validation — quantitative test of biphasic structure recovery against Liao et al. 2025 Figs 5b and 6c.

Figure 8. Model architecture schematic.